\documentclass[final,5p,times,twocolumn]{elsarticle}
\usepackage{graphicx}
\usepackage{lineno}
\usepackage{color}
\usepackage{ulem}

\journal{Astroparticle Physics}

\begin{document}

\begin{frontmatter}

\title{Search for event bursts in XMASS-I associated with gravitational-wave events}

\author[ICRR,IPMU]{K.~Abe}
\author[ICRR,IPMU]{K.~Hiraide}
\author[ICRR,IPMU]{K.~Ichimura\fnref{RCNS}}
\author[ICRR,IPMU]{Y.~Kishimoto\fnref{RCNS}}
\author[ICRR,IPMU]{K.~Kobayashi\fnref{Waseda}}
\author[ICRR]{M.~Kobayashi\fnref{Columbia}}
\author[ICRR,IPMU]{S.~Moriyama}
\author[ICRR,IPMU]{M.~Nakahata}
\author[ICRR,IPMU]{H.~Ogawa\fnref{Nihon}}
\author[ICRR]{K.~Sato\fnref{ISEE2}}
\author[ICRR,IPMU]{H.~Sekiya}
\author[ICRR]{T.~Suzuki}
\author[ICRR,IPMU]{A.~Takeda}
\author[ICRR]{S.~Tasaka}
\author[ICRR,IPMU]{M.~Yamashita\fnref{ISEE2}}
\author[ICRR,IPMU]{B.~S.~Yang\fnref{IBS2}}
\author[IBS]{N.~Y.~Kim}
\author[IBS]{Y.~D.~Kim}
\author[ISEE,KMI]{Y.~Itow}
\author[ISEE]{K.~Kanzawa}
\author[ISEE]{K.~Masuda}
\author[IPMU]{K.~Martens}
\author[IPMU]{Y.~Suzuki}
\author[IPMU]{B.~D.~Xu\fnref{Tsinghua}}
\author[Kobe]{K.~Miuchi}
\author[Kobe]{N.~Oka}
\author[Kobe,IPMU]{Y.~Takeuchi}
\author[KRISS,IBS]{Y.~H.~Kim}
\author[KRISS]{K.~B.~Lee}
\author[KRISS]{M.~K.~Lee}
\author[Miyagi]{Y.~Fukuda}
\author[Tokai1]{M.~Miyasaka}
\author[Tokai1]{K.~Nishijima}
\author[Tokushima]{K.~Fushimi}
\author[Tokushima]{G.~Kanzaki}
\author[YNU1]{S.~Nakamura}

\address[ICRR]{Kamioka Observatory, Institute for Cosmic Ray Research, the University of Tokyo, Higashi-Mozumi, Kamioka, Hida, Gifu, 506-1205, Japan}
\address[IBS]{Center for Underground Physics, Institute for Basic Science, 70 Yuseong-daero 1689-gil, Yuseong-gu, Daejeon, 305-811, South Korea}
\address[ISEE]{Institute for Space-Earth Environmental Research, Nagoya University, Nagoya, Aichi 464-8601, Japan}
\address[IPMU]{Kavli Institute for the Physics and Mathematics of the Universe (WPI), the University of Tokyo, Kashiwa, Chiba, 277-8582, Japan}
\address[KMI]{Kobayashi-Maskawa Institute for the Origin of Particles and the Universe, Nagoya University, Furo-cho, Chikusa-ku, Nagoya, Aichi, 464-8602, Japan}
\address[Kobe]{Department of Physics, Kobe University, Kobe, Hyogo 657-8501, Japan}
\address[KRISS]{Korea Research Institute of Standards and Science, Daejeon 305-340, South Korea}
\address[Miyagi]{Department of Physics, Miyagi University of Education, Sendai, Miyagi 980-0845, Japan}
\address[Tokai1]{Department of Physics, Tokai University, Hiratsuka, Kanagawa 259-1292, Japan}
\address[Tokushima]{Department of Physics, Tokushima University, 2-1 Minami Josanjimacho Tokushima city, Tokushima, 770-8506, Japan}
\address[YNU1]{Department of Physics, Faculty of Engineering, Yokohama National University, Yokohama, Kanagawa 240-8501, Japan}

\address{\rm\normalsize XMASS Collaboration$^*$}
\cortext[cor1]{{\it E-mail address:} xmass.publications14@km.icrr.u-tokyo.ac.jp .}

\fntext[RCNS]{Now at Research Center for Neutrino Scientce, Tohoku Univeristy, Sendai 980-8578, Japan.}
\fntext[Waseda]{Now at Waseda Research Institute for Science and Engineering, Waseda University, 3-4-1 Okubo, Shinjuku, Tokyo 169-8555, Japan.}
\fntext[Columbia]{Now at Physics Department, Columbia University, New York, NY 10027, USA.}
\fntext[Nihon]{Now at Department of Physics, College of Science and Technology, Nihon University, Kanda, Chiyoda-ku, Tokyo 101-8308, Japan.}
\fntext[ISEE2]{Now at Institute for Space-Earth Environmental Research, Nagoya University, Nagoya, Aichi 464-8601, Japan.}
\fntext[IBS2]{Now at Center for Axion and Precision Physics Research, Institute for Basic Science, Daejeon 34051, South Korea.}
\fntext[Tsinghua]{Department of Engineering Physics, Tsinghua University, Haidian District, Beijing 100084, China.}

\begin{abstract}
We performed a search for event bursts in the XMASS-I detector associated with
11 gravitational-wave events detected during LIGO/Virgo's O1 and O2 periods.
Simple and loose cuts were applied to the data collected in the full 832~kg xenon volume
around the detection time of each gravitational-wave event.
The data were divided into four energy regions ranging from keV to MeV.
Without assuming any particular burst models,
we looked for event bursts in sliding windows with various time width from 0.02 to 10~s.
The search was conducted in a time window between $-400$ and $+10,000$~s from each gravitational-wave event.
For the binary neutron star merger GW170817, no significant event burst was observed in the XMASS-I detector
and we set 90\% confidence level upper limits on neutrino fluence
for the sum of all the neutrino flavors via coherent elastic neutrino-nucleus scattering.
The obtained upper limit was (1.3--2.1)$\times 10^{11}$~cm$^{-2}$ under the assumption of a Fermi-Dirac spectrum
with average neutrino energy of 20~MeV.
The neutrino fluence limits for mono-energetic neutrinos in the energy range between
14 and 100~MeV were also calculated.
Among the other 10 gravitational wave events detected as the binary black hole mergers,
a burst candidate with a 3.0$\sigma$ significance was found at 1801.95--1803.95~s in the analysis for GW151012.
However, no significant deviation from the background in the reconstructed energy and position distributions was found.
Considering the additional look-elsewhere effect of analyzing the 11 GW events,
the significance of finding such a burst candidate associated with any of them is 2.1$\sigma$.
\end{abstract}

\begin{keyword}
event burst \sep gravitational wave \sep neutrino \sep astroparticle \sep liquid xenon


\end{keyword}

\end{frontmatter}


\section{Introduction}
In 2015, the gravitational-wave (GW) signal from a binary black-hole merger
was firstly detected by the Advanced LIGO experiment~\cite{Abbott:2016blz}.
During LIGO/Virgo's observing periods O1 (September 2015--January 2016) and O2 (November 2016--August 2017),
10 binary black-hole mergers and a binary neutron-star merger were observed~\cite{LIGOScientific:2018mvr}.
Moreover, the electromagnetic counterparts were detected, for the first time,
associated with the GW event from the binary neutron-star merger named GW170817~\cite{TheLIGOScientific:2017qsa};
a short gamma-ray burst was detected $\sim$1.7~s after the GW event, and 
subsequent ultraviolet, optical, and infrared emissions were also observed~\cite{GBM:2017lvd}.
Thus, a new era of the field of GW astronomy with multi-messenger observations has begun.

\begin{table*}[tb]
 \caption{List of the GW events during the whole XMASS-I data taking period.
 The data-taking situation of the XMASS-I detector in a time window between $-400$ and $+10,000$~s
 from each GW event ($t_{\rm GW}$) is also noted.}
 \label{table:dataset}
 \begin{center}
  \begin{tabular}{lccc}
    \hline \hline
    GW event & GW detection time $t_{\rm GW}$ (UTC) & Data-taking situation of XMASS-I \\
    \hline
    GW150914  & Sep. 14, 2015 09:50:45 & Continuous data-taking \\
    GW151012  & Oct. 12, 2015 09:54:43 & No data in $1,183 <t-t_{\rm GW}<1,583$~s due to run change \\
    GW151226  & Dec. 26, 2015 03:38:53 & No data in $4,191 <t-t_{\rm GW}< 4,388$~s due to run change \\
    GW170104  & Jan. 04, 2017 10:11:58 & No data in $196 <t-t_{\rm GW}< 275$~s due to run change \\
    GW170608  & Jun. 08, 2017 02:01:16 & No data in $t-t_{\rm GW}> 6,339$~s due to detector calibration \\
    GW170729  & Jul. 29, 2017 18:56:29 & Continuous data-taking \\
    GW170809  & Aug. 09, 2017 08:28:21 & Continuous data-taking \\
    GW170814  & Aug. 14, 2017 10:30:43 & Continuous data-taking \\
    GW170817  & Aug. 17, 2017 12:41:04 & Continuous data-taking \\
    GW170818  & Aug. 18, 2017 02:25:09 & Continuous data-taking \\      
    GW170823  & Aug. 23, 2017 13:13:58 & Continuous data-taking \\   
    \hline \hline
  \end{tabular}
 \end{center}
\end{table*}

The follow-up searches for neutrino events associated with these GW events
have also been conducted by gigantic neutrino detectors all over the world,
however, no significant neutrino signal has been observed yet~\cite{Adrian-Martinez:2016xgn,Aab:2016ras,Gando:2016zhq,Abe:2016jwn,Agostini:2017pfa,ANTARES:2017bia,Abe:2018mic,Acero:2020duu}.
The neutrino follow-up searches are of interest because, for instance,
there are some theoretical predictions of emission of neutrinos
with energy of a few tens MeV~\cite{Sekiguchi:2011zd,Kyutoku:2017wnb},
and much higher-energy neutrinos~\cite{Kimura:2017kan,Fang:2017tla} are expected
from binary neutron-star mergers.
The expected range for detecting several neutrinos from a binary neutron star merger is, however,
$<10$~Mpc with a future megaton-scale water Cherenkov detector~\cite{Sekiguchi:2011zd}.
There also exist scenarios of production of short gamma-ray bursts~\cite{Perna:2016jqh}
or ultrahigh-energy neutrinos~\cite{Kotera:2016dmp} from binary black-hole mergers. 

XMASS-I is a large single-phase liquid xenon (LXe) detector located underground (2700~m water equivalent)
at the Kamioka Observatory in Japan~\cite{xmass-detector}.
It is a multi-purpose detector suitable for detecting particles with energy deposition
in the wide energy range from keV to MeV under an ultra-low background environment.
The XMASS-I detector accumulated data with a stable condition continuing from November 2013 until February 2019,
resulting in the entire data set with a total live time of 4.4 years.
Using the XMASS-I data, various searches for astroparticles such as
dark matter particles~\cite{xmass-modulation2,xmass-super-wimps2,xmass-fv-wimps,xmass-sub-gev},
solar axions~\cite{xmass-solar-axion}, and solar Kaluza-Klein axions~\cite{xmass-kk-axion}
have been performed.
Furthermore, the possibility to detect galactic supernova neutrinos
via coherent elastic neutrino-nucleus scattering (CEvNS) has been studied~\cite{xmass-supernova}.

In this paper, we report results from a search for event bursts in the XMASS-I detector
associated with the 11 GW events detected during LIGO/Virgo's O1 and O2 periods.

\section{XMASS-I detector}
The XMASS-I detector holds an active target of 832~kg of LXe inside a pentakis-dodecahedral copper structure
that hosts 642 inward-looking 2-inch Hamamatsu R10789 photomultiplier tubes (PMTs)
on its approximately spherical inner surface at a radius of about 40~cm.
The photocathode coverage of the inner surface is 62.4\%.
The LXe detector is placed at the center of a cylindrical water Cherenkov detector.
The outer detector, which is 11~m in height and 10~m in diameter,
is equipped with 72 20-inch Hamamatsu H3600 PMTs.
This detector acts as an active veto counter for cosmic-ray muons as well as
a passive shield against neutrons and $\gamma$-rays from the surrounding rocks.

Data acquisition is triggered if at least four inner-detector PMTs record
a signal within 200~ns
or if at least eight outer-detector PMTs register a signal within 200~ns.
A 50~MHz clock is used to measure the time difference between triggers.
One-pulse-per-second (1PPS) signals from the global positioning system (GPS) are fed as triggers
for precise time stamping.
The GPS 1PPS triggers are also used to flash the LED for the PMT gain monitoring.

The gains of the PMTs are monitored weekly using a blue LED embedded in the inner surface of the detector. 
The scintillation yield response is traced with a $^{57}$Co source~\cite{xmass-source}
inserted along the central vertical axis of the detector every week or two. 
Through measurements with the $^{57}$Co source at the center of the detector volume, 
the photoelectron (PE) yield was determined to be $\sim$15~PE/keV for 122~keV $\gamma$-rays.
The nonlinear response of the scintillation yield for electron-mediated events in the detector
was calibrated over the energy range from 5.9~keV to 2614~keV
with $^{55}$Fe, $^{241}$Am, $^{109}$Cd, $^{57}$Co, $^{137}$Cs, $^{60}$Co, and $^{232}$Th sources.
In this paper, the energy above 2614~keV is extrapolated using the energy scale derived by
the $^{232}$Th calibration.
Hereinafter, this calibrated energy is represented as keV$_{\rm ee}$
where the subscript stands for the electron-equivalent energy.

The timing offsets for the PMT channels owing to the differences in their cable lengths and
the electronic responses were also traced by the $^{57}$Co calibration.

\section{Data set and event selection}
Table~\ref{table:dataset} shows a list of the GW events detected during LIGO/Virgo's O1 and O2 periods and
the data-taking situation of the XMASS-I detector around the detection time of each GW event.
The event burst search is conducted in a time window between $-400$ and $+10,000$~s from each GW event.
This search window is motivated by two reasons. The time window within $\pm$400~s from each GW event is considered
in the search for neutrinos as described in Sec.~\ref{sec:neutrino}. In addition, the extended time window up to 10,000 s is considered
in the model independent search because more massive particles, axion-like particles for instance, might arrive later timing.
For the GW151012, GW151226, and GW170104 events, there exists
a-few-minute dead time due to run change.
For the GW170608, data-taking stopped $6,339$~s after the GW event for the detector calibration.
Otherwise, data were taken continuously during the search window.

We use the full 832~kg of xenon as an active target in this analysis.
In order to perform a search for event bursts with a minimal bias,
only simple and loose cuts are applied.
Events with four or more hits in the inner-detector without an associated
outer-detector trigger are initially selected.
We have then applied four selection cuts that mostly remove obvious backgrounds.
To remove events caused by after-pulses in the PMTs following bright events,
one requires that the standard deviation of the inner-detector hit timing distribution
is less than 100~ns, and that the time elapsed since the previous inner-detector event
($dT_{\rm pre}$) is at least 200~$\mu$s.
To remove events due to Cherenkov light emission by the $\beta$-rays from $^{40}$K in the PMT photocathode,
events in which more than 80\% of the PMT hits arrive in the first 20~ns are discarded.
For this analysis, the $dT_{\rm pre}$ and Cherenkov cuts are loosen compared to other analyses in XMASS-I.
The detection efficiency after those cuts will be described later.

\begin{figure}[tp]
  \begin{center}
    \includegraphics[keepaspectratio=true,width=95mm]{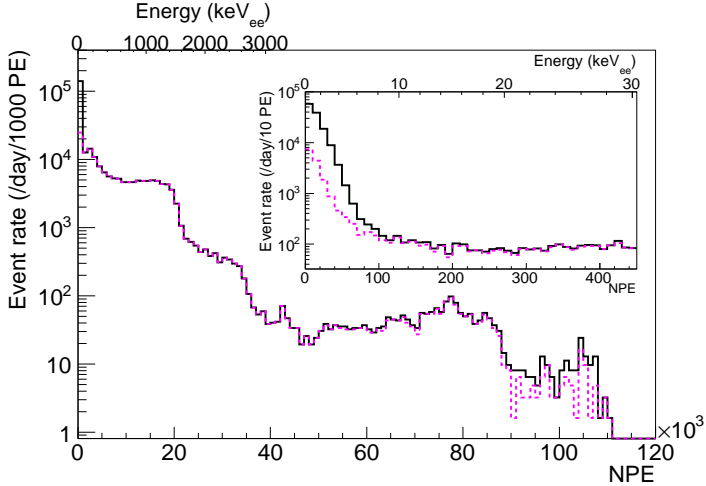}
  \end{center}
  \caption{Observed number of PE spectra in the pre-search time window for the GW170817 event
  before (black solid) and after (magenta dashed) the event selection.
  The corresponding keV$_{\rm ee}$ energy scale is shown at the top of the figure.}
  \label{fig:data-spectrum.eps}
\end{figure}

Figure~\ref{fig:data-spectrum.eps} shows the observed number of PE (NPE) spectra in the pre-search time window for the GW170817 event
before and after the event selection.
The pre-search window is defined as a time window between $-20,000$ and $-400$ s from each GW event to estimate the background rate. 
The cuts mainly remove events below 100~PE as well as a fraction of $\alpha$-ray events from the $^{222}$Rn daughter, $^{214}$Po.
Then, the data are divided into four NPE ranges: $<$450 (referred to as Low-E),
450--4500 (Middle-E), 4500--45000 (High-E), and $>$45000 (Very High-E, or V.~H.~E. hereinafter)
corresponding to energy ranges of approximately $<$30, 30--300, 300--3500, and $>$3500~keV$_{\rm ee}$, respectively.
The average event rate in these four energy ranges are estimated from the pre-search window,
to be 0.223$\pm$0.004, 0.559$\pm$0.006, 0.987$\pm$0.008, and 0.023$\pm$0.001~Hz, respectively.

\section{Model independent event burst search}
\subsection{Results for the GW170817 binary neutron star merger}

\begin{figure}[tp]
  \begin{center}
    \includegraphics[keepaspectratio=true,width=95mm]{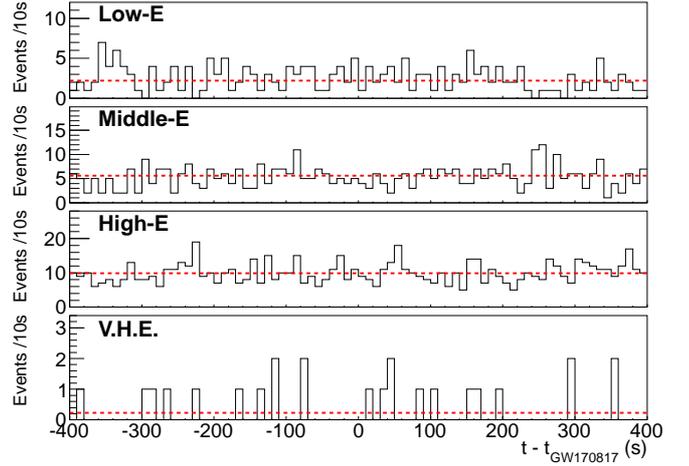}
  \end{center}
  \caption{Event rate history within $\pm$400~s from GW170817 (August 17, 2017, 12:41:04~UTC).
  From top to bottom, the Low-E, Middle-E, High-E, and V. H. E. samples are shown.
  The horizontal dashed lines correspond to the average background event rate estimated
  from the pre-search window.}
  \label{fig:rate-400s.eps}
\end{figure}

\begin{figure}[tp]
  \begin{center}
    \includegraphics[keepaspectratio=true,width=95mm]{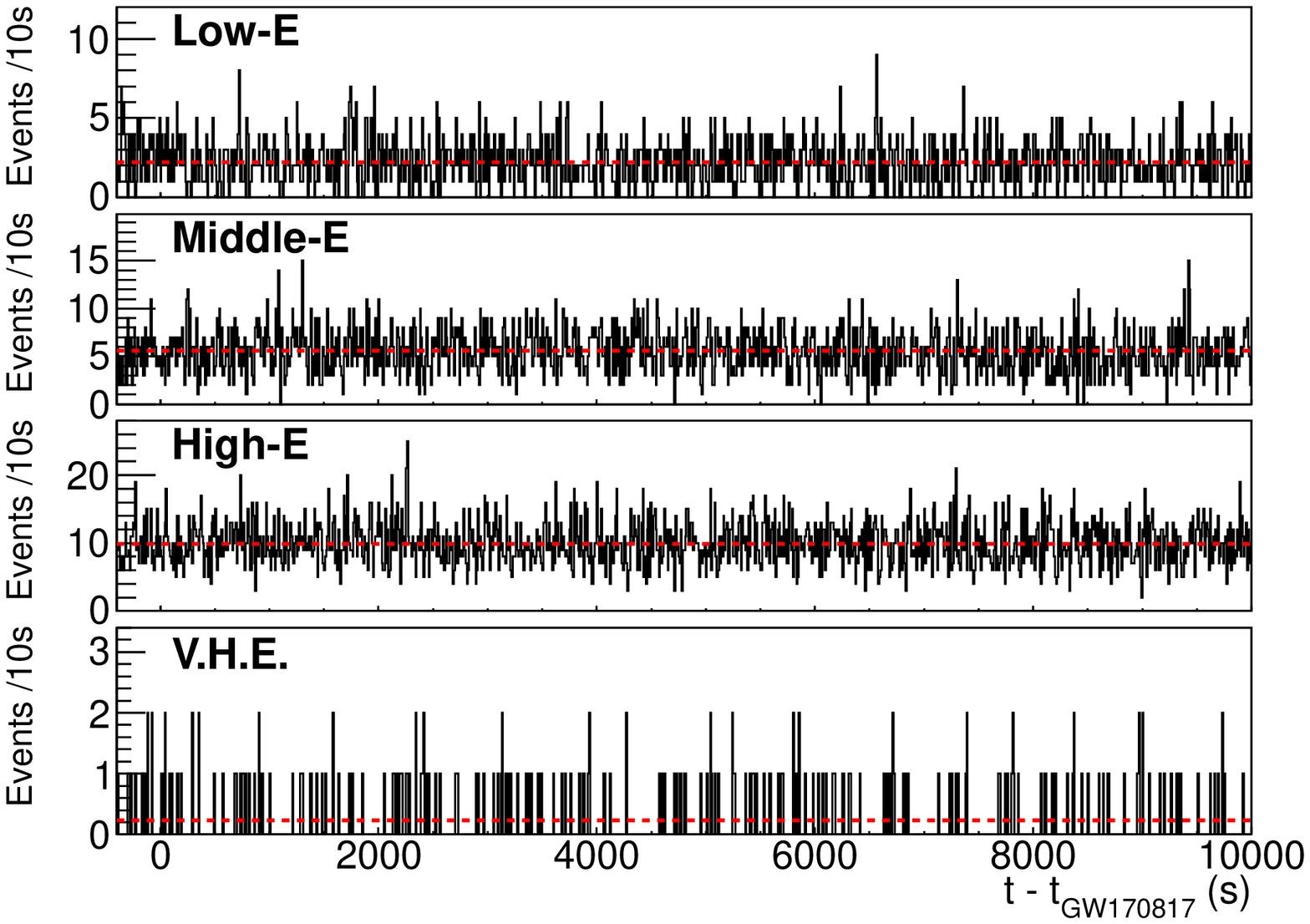}
  \end{center}
  \caption{Event rate history in a wider time range up to $+10,000$~s from GW170817 (August 17, 2017 12:41:04~UTC).
  From top to bottom, the Low-E, Middle-E, High-E, and V. H. E. samples are shown.
  The horizontal dashed lines correspond to the average background event rate estimated
  from the pre-search window.}
  \label{fig:rate-10000s.eps}
\end{figure}

\begin{figure}[tp]
  \begin{center}
    \includegraphics[keepaspectratio=true,width=95mm]{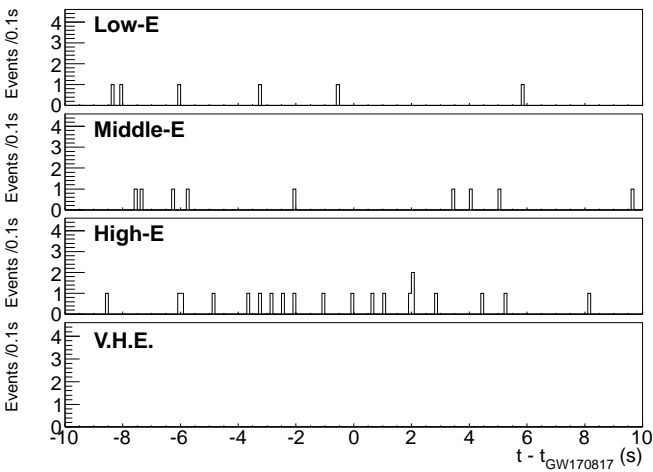}
  \end{center}
  \caption{Event rate history within $\pm$10~s from GW170817 (August 17, 2017 12:41:04~UTC).
  From top to bottom, the Low-E, Middle-E, High-E, and V. H. E. samples are shown.}
  \label{fig:rate-ontime.eps}
\end{figure}

Figures~\ref{fig:rate-400s.eps} shows the event rate history within $\pm$400~s from GW170817 (August 17, 2017 12:41:04~UTC).
The event rate history in a wider time range up to $+10,000$~s from the GW event is shown in Fig.~\ref{fig:rate-10000s.eps}.
The average background event rates estimated from the pre-search window are shown as horizontal dashed lines.
A close-up event rate history within $\pm$10~s from GW170817 is also shown in Fig.~\ref{fig:rate-ontime.eps}.

\begin{figure}[tp]
  \begin{center}
    \includegraphics[keepaspectratio=true,width=90mm]{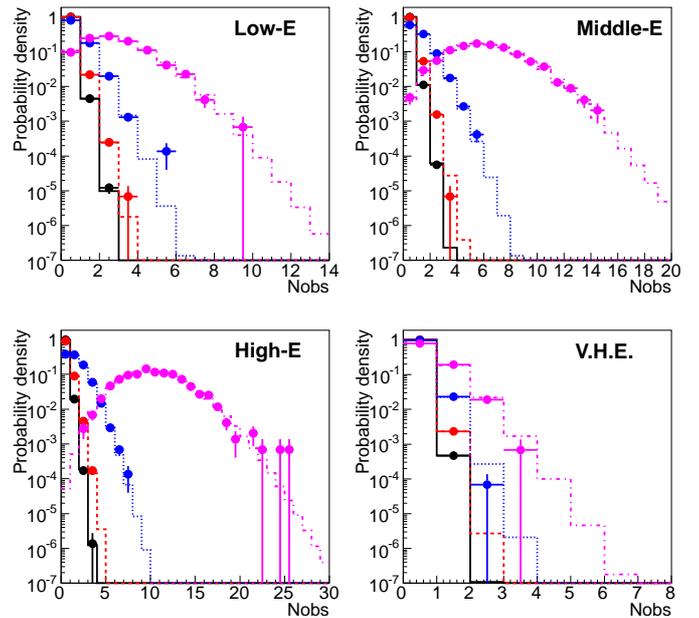}
  \end{center}
  \caption{Distributions of the observed number of events ($N_{\rm obs}$) in time windows of
  width 0.02~s (black solid), 0.1~s (red dashed), 1~s (blue dotted) and 10~s (magenta dash-dotted) in each energy range.
  The points are the data from the pre-search window and the histograms are obtained from the simulation described in the text.}
  \label{fig:nevents_per_bin.eps}
\end{figure}

To search for event bursts without assuming any particular burst model,
the number of events in a sliding time window is scanned for each energy range.
The coincidence time window is slid at a step of 0.01~s and
various width of the window ($t_{\rm width}$) are tested: 0.02, 0.04, 0.1, 0.2, 0.4, 1, 2, 4, and 10 s.
The range of the width of the sliding window is determined referring to the neutrino emission model
discussed in ref.~\cite{Sekiguchi:2011zd}; the neutrino emission peak continues 10-100 ms with a long cooling
time of 2-3 s is predicted.
For each energy region, a test statistics (TS) is constructed under the null hypothesis as
\begin{equation}
  {\rm TS} = 2 \ln \left[ \left( \frac{T_{\rm search}}{t_{\rm width}} \right) \times P(\mu_{\rm bg}|N_{\rm obs}) \right] \ ,
\end{equation}
where $P(\mu|N)= \mu^{N} e^{-\mu} / N!$ is the Poisson probability, and
$N_{\rm obs}$ and $\mu_{\rm bg}$ are the observed number of events and
the expected number of background events in a time window.
The factor $T_{\rm search}/t_{\rm width}$ corrects for the look-elsewhere effect due to
choosing a time window of width $t_{\rm width}$ from the whole search window $T_{\rm search}$.

At each timing $t-t_{\rm GW}$, we first minimize the test statistics as a function of $t_{\rm width}$.
Then, we scan the test statistics as a function of the timing $t-t_{\rm GW}$ to find minimums
as possible burst candidates. 

To calculate a global significance of each candidate, we estimate the probability distribution
of the test statistics under the null hypothesis by performing the same analysis on 100,000 dummy data sets.
In each dummy data set, events are randomly generated based on the average event rate estimated
in the pre-search window for each energy region.
Possible time-correlated backgrounds due to short-time consecutive decays of radioisotopes in the detector material are also considered.
$^{222}$Rn in the LXe decays through the $^{222}$Rn ($T_{1/2}$=3.82 d)--$^{218}$Po ($T_{1/2}$=3.10~min)--$^{214}$Pb ($T_{1/2}$=26.8 min)
chain, and 5.49 and 6.00~MeV $\alpha$-ray events could occur in the same time window in the V.H.E. energy range.
This background is taken into account in the dummy data generation based on the measured $^{222}$Rn activity of $\sim$8~mBq in the active LXe volume~\cite{xmass-fv-wimps}.
The contributions from short-time consecutive decays in the $^{238}$U and $^{232}$Th chains contaminated in the detector material
are turned out to be negligible.
Figure~\ref{fig:nevents_per_bin.eps} shows the distributions of the observed number of events
in coincidence windows with various widths for the data of the pre-search window overlaid with the simulation.
The simulated distributions well reproduce the data of the pre-search window.

Finally, the look-elsewhere effect due to the search using 4 energy windows is accounted for.
For small $p$-values, this correction can be made by multiplying the $p$-value by the number of energy ranges, that is 4.

As the result of the model independent search,
no coincidence time window with a global significance of more than 3$\sigma$ was found
in the time range between $-400$ and $+10,000$~s from the GW170817 event.

\subsection{Results for the binary black hole mergers}

\begin{figure}[tp]
  \begin{center}
    \includegraphics[keepaspectratio=true,width=90mm]{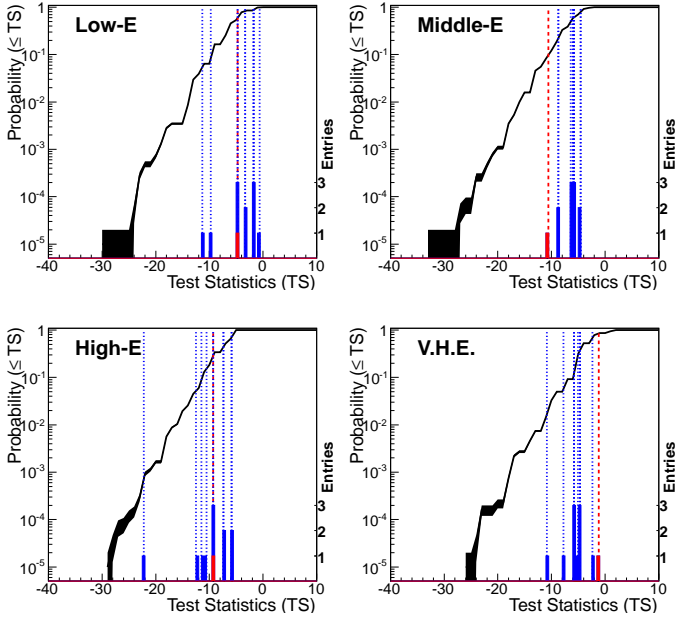}
  \end{center}
  \caption{Distributions of the test statistics (TS) observed in each energy range
  for 10 binary black-hole mergers (blue dotted) and a binary neutron-star merger (red dashed).
  The black curve shows the probability of observing smaller TS value obtained
  from the 100,000 dummy data sets.}
  \label{fig:ts-cdf.eps}
\end{figure}

\begin{figure}[tp]
  \begin{center}
    \includegraphics[keepaspectratio=true,width=95mm]{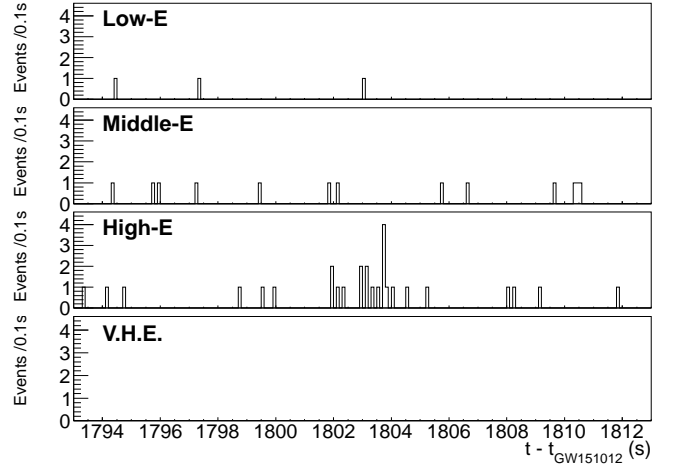}
  \end{center}
  \caption{Event rate history within $\pm$10~s around the burst candidate found centered at $t-t_{\rm GW} =1802.95$~s from GW151012
   (October 12, 2015 09:54:43~UTC). From top to bottom, the Low-E, Middle-E, High-E, and V. H. E. samples are shown.}
  \label{fig:GW151012-rate.eps}
\end{figure}

\begin{figure}[tp]
  \begin{center}
    \includegraphics[keepaspectratio=true,width=95mm]{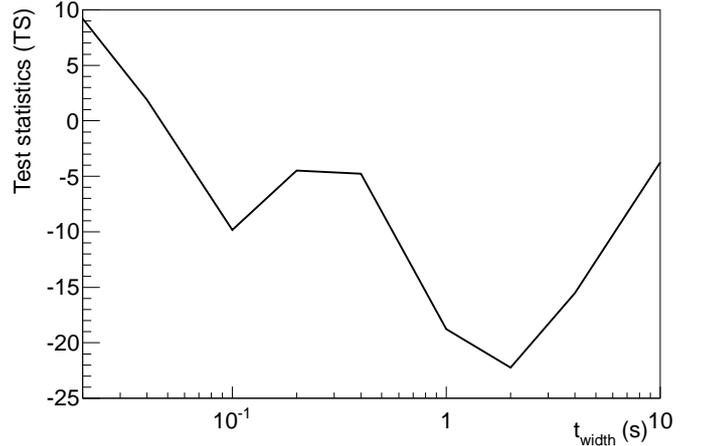}
  \end{center}
  \caption{Test statistics (TS) as a function of coincidence time width $t_{\rm width}$
  in the High-E energy range around $t-t_{\rm GW}=1802.95$~s from GW151012.
  Note that $t_{\rm width}$ is scanned discretely at 0.02, 0.04, 0.1, 0.2, 0.4, 1, 2, 4, and 10 s.}
  \label{fig:GW151012-ts-vs-twidth.eps}
\end{figure}

The same analysis is performed for other 10 GW events classified as the binary black hole mergers.
Figure~\ref{fig:ts-cdf.eps} shows the distributions of the TS observed in each energy range
for 10 binary black-hole mergers and a binary neutron-star merger.
Among them, a burst candidate with small probability of $P(\le {\rm TS}) = 7.8\times 10^{-4}$ 
was found in the High-E energy range for the GW151012 event.
As seen in Fig.~\ref{fig:GW151012-rate.eps}, 15 events are clustered within ~2 s centered at $t-t_{\rm GW} = 1802.95$~s.
Figure~\ref{fig:GW151012-ts-vs-twidth.eps} shows the test statistics as a function of
the coincidence time width in the High-E energy range around $t-t_{\rm GW151012}=1802.95$~s.
The coincidence time width of 2~s gives the lowest TS value, and hence the most significant result. 
After considering the look-elsewhere effect of the 4 energy ranges,
the global significance of this burst candidate identified in association with GW151012 is 3.0$\sigma$.
Since we perform the analysis separately on the 11 GW events, there is an additional look-elsewhere effect.
The significance of finding such a burst candidate in any of the 11 GW events is 2.1$\sigma$.

\begin{figure}[tp]
  \begin{center}
    \includegraphics[keepaspectratio=true,width=95mm]{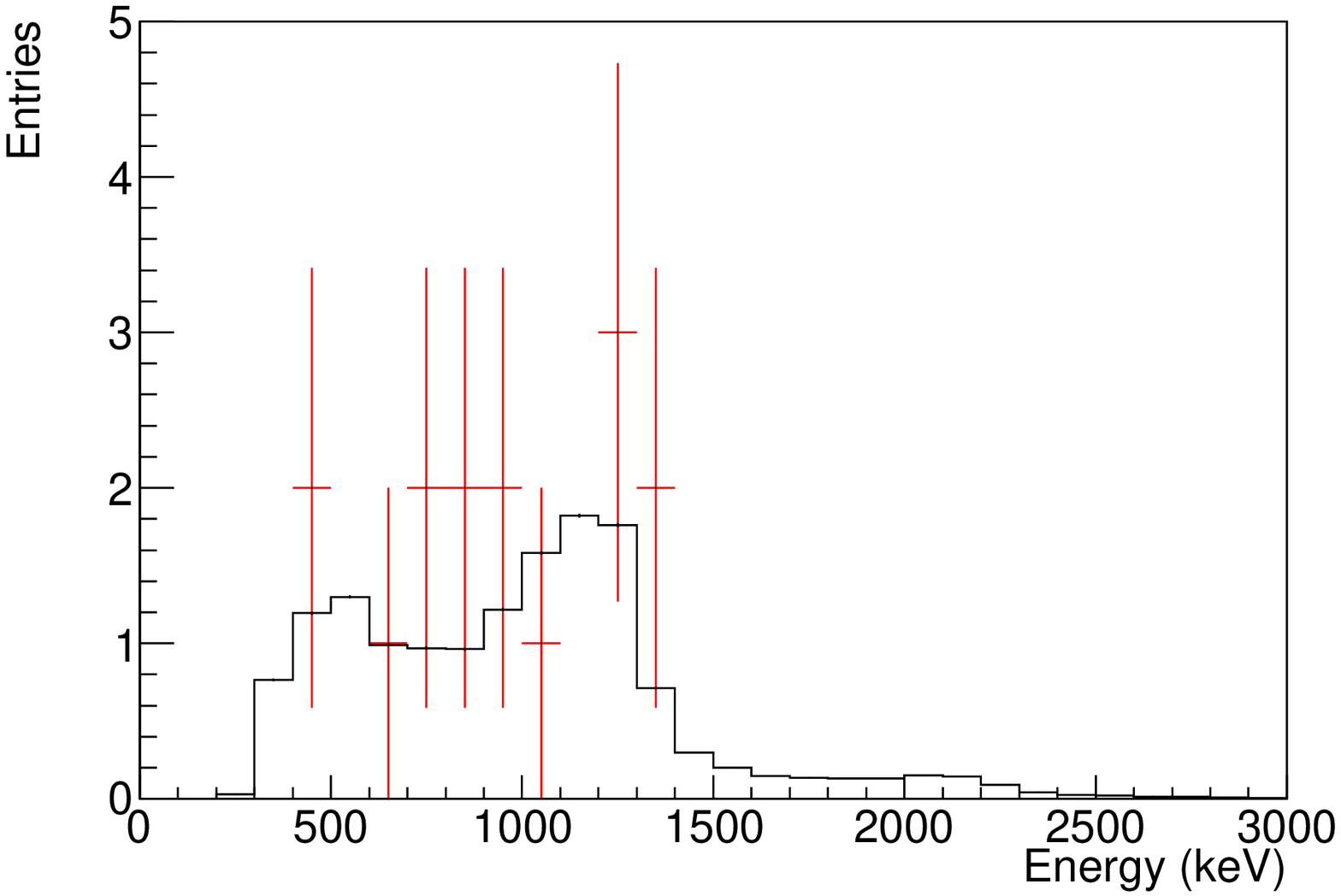}
    \includegraphics[keepaspectratio=true,width=95mm]{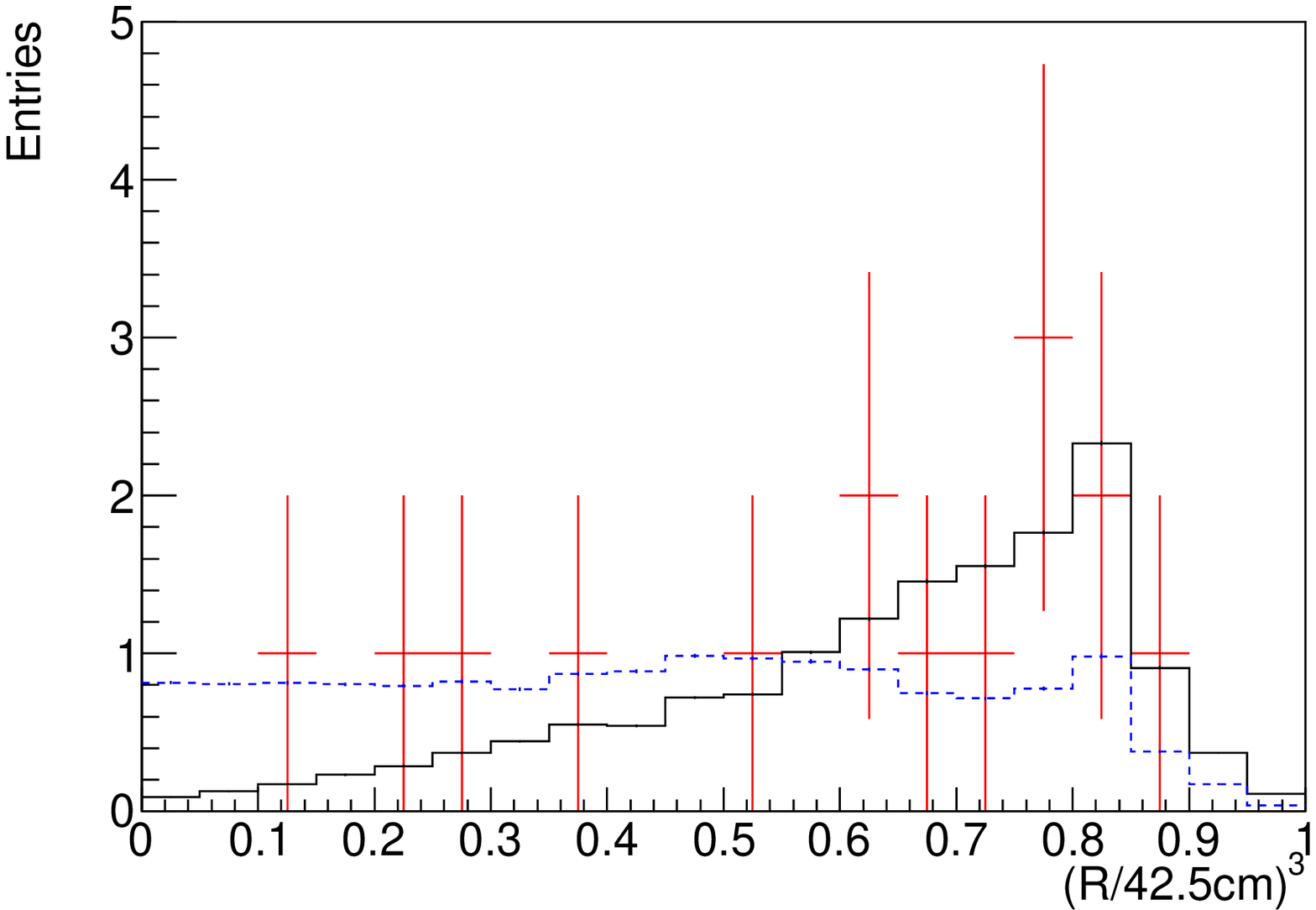}
  \end{center}
  \caption{Reconstructed energy (top) and radial position (bottom) of the 15 events in the burst candidate
  overlaid with the distributions for background estimated using the pre-search window (black solid).
  The simulated radial distribution for the uniformly distributed events (blue dashed) is also shown for comparison.
  The distributions for background and the uniformly distributed events are normalized to the observed number of events.}
  \label{fig:GW151012-energy-r3.eps}
\end{figure}

The energy and vertex position of those events are reconstructed based on a maximum-likelihood evaluation of the observed
NPE of all the PMTs~\cite{xmass-detector}.
Note that the XMASS-I detector observes the LXe scintillation light, and therefore provides no directional information on detected particles.
Figure~\ref{fig:GW151012-energy-r3.eps} shows the reconstructed energy and radial position distributions of those 15 events
overlaid with the ones for the background estimated using the pre-search window.
The $p$-value of the Kolmogorov-Smirnov (KS) test for the energy and radial position
distributions between those 15 events and the background are 0.87 and 0.96, respectively.
Therefore, no significant deviation from the background distributions is found.

To investigate whether the radial distribution of the burst candidate is consistent with
that of potential signal uniformly distributed inside the detector,
we overlay the radial distribution for uniformly distributed electron events 
(blue dashed in Fig.~\ref{fig:GW151012-energy-r3.eps}) using our GEANT4-based detector simulation.
Since we do not have any signal model predicting the energy spectrum,
a flat energy spectrum up to 3~MeV is assumed.
A dip at $(R/42.5{\rm cm})^3 \sim 0.7$ in the reconstructed radial distribution for the simulation is
due to a slight position dependent bias in the radial direction. Note that the edge of
the pentakis-dodecahedral detector locates between $(R/42.5{\rm cm})^3 =0.83$ and 1.
The $p$-value of the KS test for the uniformly distributed events is 0.055.

We concluded that the radial distribution of the burst candidate is compatible with
both that of the background and that of the uniform distribution 
since the obtained p-values of the KS test under both hypotheses are above 0.05 (or within 2$\sigma$).
We also estimated the frequency that 15 events in the High-E energy region are observed within 2~s
due to statistical fluctuation during the entire data-taking period of XMASS-I to be 0.21~yr$^{-1}$,
based on the average event rate in the pre-search window.

\section{Constraints on neutrino fluence for GW170817}
\label{sec:neutrino}
For the GW170817 event, we also derive constraints on neutrino fluence
for the sum of all neutrino flavors via CEvNS 
under the assumption of two types of neutrino energy distributions:
a Fermi-Dirac spectrum with average neutrino energy of 20~MeV
and mono-energetic neutrinos in the range between 14 and 100~MeV.
The relative difference in propagation time for GW and such neutrinos and GWs
from the source, for example, at a 40~Mpc distance is expected to be $<O(1)$~s.
For this neutrino signal search, we use the $\pm$400~s search window,
which is a similar size of the search window as the one used in searches by other neutrino
detectors~\cite{Adrian-Martinez:2016xgn,Aab:2016ras,Gando:2016zhq,Abe:2016jwn,Agostini:2017pfa,ANTARES:2017bia,Abe:2018mic,Acero:2020duu}.

The Fermi-Dirac energy distribution is expressed as
\begin{equation}
    f(E_\nu) = \frac{C}{(k_{\rm B} T)^3} \frac{E_\nu^2}{e^{E_\nu/k_{\rm B} T} +1} ,
\end{equation}
where
\begin{equation}
    C = \left( \int_0^\infty \frac{x^2}{e^x+1} {\rm d} x \right)^{-1} = \frac{2}{3 \zeta (3)}
\end{equation}
is the normalization factor, $k_{\rm B}$ is the Boltzmann constant, $T$ is temperature,
and the average energy of neutrinos is given by $\langle E_\nu \rangle \sim 3.15T$.

The differential cross section of CEvNS is
\begin{equation}
    \frac{\rm{d} \sigma}{\rm{d}E_{\rm nr}} (E_\nu, E_{\rm nr})
    = \frac{G_F^2 M}{2\pi} G_V^2 \left [ 1 + \left( 1-\frac{E_{\rm nr}}{E_\nu} \right )^2-\frac{ME_{\rm nr}}{E_\nu^2}  \right ] ,
\end{equation}
where $G_F$ is the Fermi constant, $M$ is the target nuclear mass, $E_{\rm nr}$ is the nuclear recoil energy, and
\begin{equation}
    G_V = \left [ \left ( \frac{1}{2} -2\sin^2 \theta_W \right ) Z -\frac{1}{2}N \right ] F(q^2) .
\end{equation}
Here, $\theta_{\rm W}$ is the weak mixing angle,
$Z$ and $N$ are the numbers of protons and neutrons in the nucleus,
and $F(q^2)$ is the nuclear form factor, respectively.
More detail for the calculation of the CEvNS interaction in XMASS-I can be found elsewhere~\cite{xmass-supernova}.

\begin{figure}[tbp]
  \begin{center}
    \includegraphics[keepaspectratio=true,width=95mm]{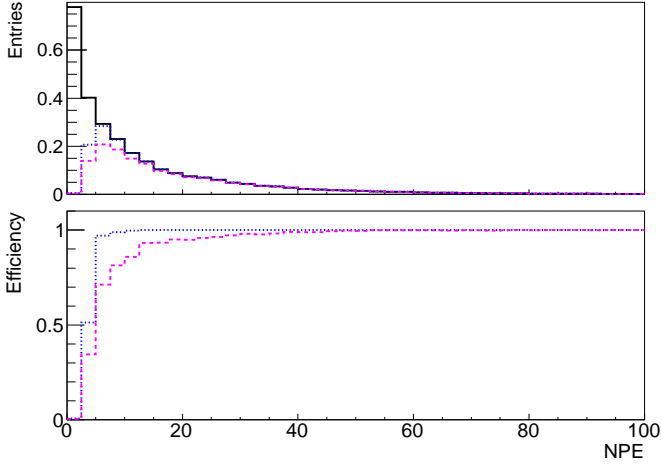}
  \end{center}
  \caption{Simulated NPE spectra (top) and detection efficiency as a function of NPE (bottom)
  for the Fermi-Dirac spectrum with  $\langle E_\nu \rangle$=20~MeV.
  The black solid, the blue dotted, and the magenta dashed histograms are without any cut, after the inner-detector trigger selection,
  and after the Cherenkov cut, respectively.}
  \label{fig:fermi-dirac-eff.eps}
\end{figure}

Figure~\ref{fig:fermi-dirac-eff.eps} shows the simulated NPE spectra and detection efficiency
as a function of NPE for the Fermi-Dirac spectrum with  $\langle E_\nu \rangle$=20~MeV.
The simulated neutrino events concentrate at low energy, and hence the Low-E sample (below 450~PE)
is used to derive constraints on neutrino fluence.
The detection efficiency crosses 50\% at 4.5~PE which corresponds to 3.8~keV$_{\rm nr}$.

No significant event burst is observed in the Low-E sample as described in the previous section,
the 90\% confidence level (CL) upper limit on neutrino fluence ($\Phi_{90}$) is calculated by
\begin{equation}
    \Phi_{90} = \frac{N_{90}}
    {N_T \int \int f(E_{\nu}) \frac{{\rm d} \sigma}{{\rm d} E_{\rm nr}} (E_\nu, E_{\rm nr}) \epsilon (E_{\rm nr}) \rm{d}E_\nu \rm{d}E_{nr}}
\end{equation}
where $N_T$ is the number of target nuclei, $\epsilon (E_{\rm nr})$ is the detection efficiency as a function of recoil energy
and is estimated using our detector simulation.
$N_{90}$ is the 90\% CL upper limit on the number of signal events, derived from the relation
\begin{equation}
  \frac{\int_0^{N_{90}} P(\mu_{\rm sig}+\mu_{\rm bg}|N_{\rm obs}) {\rm d} \mu_{\rm sig}}
  {\int_0^{\infty} P(\mu_{\rm sig}+\mu_{\rm bg}|N_{\rm obs}) {\rm d} \mu_{\rm sig}} = 0.9 ,
\end{equation}
where $P(\mu|N)$ is the Poisson probability,
$N_{\rm obs}$ is the observed number of events in the coincidence time window with a width $t_{\rm width}$,
and $\mu_{\rm sig}$ and $\mu_{\rm bg}$ are the average number of signal and background events, respectively.
$\mu_{\rm bg}$ is estimated based on the average event rate estimated in the pre-search window,
and we assume a uniform prior on $\mu_{sig}$.

\begin{figure}[tp]
  \begin{center}
    \includegraphics[keepaspectratio=true,width=95mm]{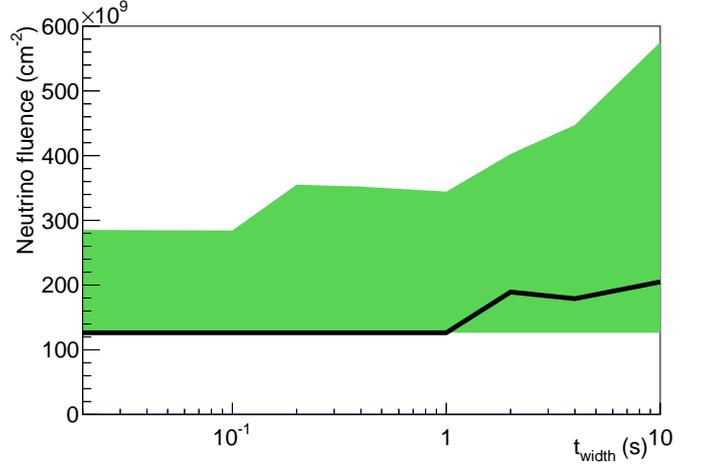}
  \end{center}
  \caption{90\% CL upper limits on neutrino fluence for GW170817 assuming the Fermi-Dirac spectrum with
  $\langle E_\nu \rangle$=20~MeV as a function of the coincidence time width $t_{\rm width}$.
  The black solid line shows the upper limit from the on-time window
  centered at $t = t_{\rm GW}$ with a width $t_{\rm width}$,
  and the green band represents the range of limits from the sliding window with a width $t_{\rm width}$
  within the $\pm 400$~s search window.
  Note that $t_{\rm width}$ is scanned discretely at 0.02, 0.04, 0.1, 0.2, 0.4, 1, 2, 4, and 10 s.}
  \label{fig:neutrino-fluence-limit-twidth.eps}
\end{figure}

Figure~\ref{fig:neutrino-fluence-limit-twidth.eps} shows the 90\% CL upper limits on
neutrino fluence for the Fermi-Dirac spectrum with $\langle E_\nu \rangle$=20~MeV
as a function of the coincidence time width $t_{\rm width}$.
The upper limit from the on-time window centered at $t = t_{\rm GW}$ with a width $t_{\rm width}$ is drawn as a line
while the range of limits from the sliding window with a width $t_{\rm width}$ within the $\pm 400$~s search window
is drawn as a band.
The obtained upper limit from the on-time window was (1.3--2.1)$\times 10^{11}$~cm$^{-2}$. 

\begin{figure}[tp]
  \begin{center}
    \includegraphics[keepaspectratio=true,width=95mm]{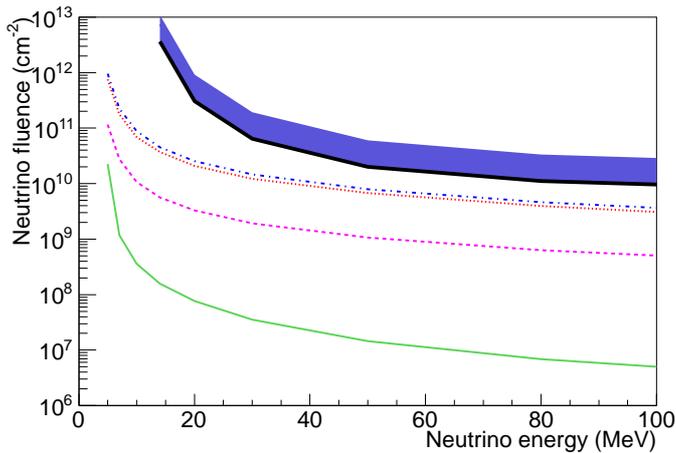}
  \end{center}
  \caption{90\% CL upper limits on mono-energetic neutrino fluence for GW170817 as a function of neutrino energy.
  The black solid line shows the upper limit from the on-time window,
  and the blue band represents the range of limits from the sliding window within the $\pm 400$~s search window.  
  Limits obtained by Super-Kamiokande~\cite{Abe:2018mic} for $\bar{\nu}_e$ (green solid), $\nu_e$ (magenta dashed),
  $\nu_{\mu,\tau}$ (red dotted), and $\bar{\nu}_{\mu,\tau}$ (blue dash-dotted) are also shown.}
  \label{fig:neutrino-fluence-limit-energy.eps}
\end{figure}

Figure~\ref{fig:neutrino-fluence-limit-energy.eps} shows our 90\% CL upper limits on fluence
for mono-energetic neutrinos as a function of neutrino energy between 14 and 100~MeV.
Limits obtained by Super-Kamiokande~\cite{Abe:2018mic} are also shown.
While their limits were derived utilizing the inverse beta decay ($\bar{\nu}_e + p \to e^+ + n$) for $\bar{\nu}_e$
and neutrino--electron elastic scattering ($\nu + e^- \to \nu + e^-$) for $\nu_e$ and $\nu_{\mu,\tau}$,
our limits are for the sum of all the neutrino flavors utilizing CEvNS.
The XMASS limit is comparable to the $\nu_{\mu,\tau}$ and $\bar{\nu}_{\mu,\tau}$ limits of Super-Kamiokande.

\section{Conclusion}
We conducted a search for event bursts in the XMASS-I detector associated with
11 GW events detected during LIGO/Virgo's O1 and O2 periods.
We used the full 832 kg of xenon as an active target.
Simple and loose cuts were applied to the data collected around the detection time of
each GW event and the data were divided into four energy regions ranging from keV to MeV.
Without assuming any particular burst model,
we looked for event bursts in sliding windows with various time width from 0.02 to 10~s and
the search was conducted in a time window between $-400$ and $+10,000$~s from each GW event.
For the binary neutron star merger GW170817, no significant event burst was observed in the XMASS-I detector,
and hence we set 90\% confidence level upper limits on neutrino fluence
for the sum of all the neutrino flavors via coherent elastic neutrino-nucleus scattering.
The obtained upper limit was (1.3--2.1)$\times 10^{11}$~cm$^{-2}$ under the assumption of the Fermi-Dirac spectrum
with the average neutrino energy of 20~MeV.
The neutrino fluence limits for mono-energetic neutrinos in the energy range between 14 and 100~MeV were also calculated.
Among the other 10 GW events detected as the binary black hole mergers,
a burst candidate with a 3.0$\sigma$ significance was found at 1801.95--1803.95~s in the analysis for GW151012.
However, the reconstructed energy and position distributions were consistent with those expected from the background.
Considering the additional look-elsewhere effect of analyzing the 11 GW events,
the significance of finding such a burst candidate associated with any of them is 2.1$\sigma$.

\section*{Acknowledgments}
We gratefully acknowledge the cooperation of the Kamioka Mining and Smelting Company.
This work was supported by the Japanese Ministry of Education, Culture, Sports, Science
and Technology, Grant-in-Aid for Scientific Research (19GS0204, 26104004, 26104007, and 26105505),
the joint research program of the Institute for Cosmic Ray Research (ICRR), the University of Tokyo,
and partially by the National Research Foundation of Korea Grant
funded by the Korean Government (NRF-2011-220-C00006).








\end{document}